# Radiomic biomarker extracted from PI-RADS 3 patients support more efficient and robust prostate cancer diagnosis: a multi-center study

Longfei Li[1,2], Rui Yang[3], Xin Chen[4], Cheng Li[5], Hairong Zheng[5], Yusong Lin[1], Zaiyi Liu[4], and Shanshan Wang[5]

[1]the Collaborative Innovation Center for Internet Healthcare, School of Information Engineering, Zhengzhou University, Zhengzhou, China, [2]Paul C. Lauterbur Research Center for Biomedical Imaging, Shenzhen Institute of Advanced Technology, Chinese Academy of Sciences, Shenzhen, China, [3]Department of Urology, Renmin Hospital of Wuhan University, Wuhan, China, [4]Department of Radiology, Guangdong Provincial People's Hospital, Guangzhou, China, [5]Paul C. Lauterbur Research Center for Biomedical Imaging, Shenzhen Institute of Advanced Technology, Chinese Academy of Sciences, Shenzhen, China

## Synopsis

Prostate Imaging Reporting and Data System (PI-RADS) based on multi-parametric MRI classifies patients into 5 categories (PI-RADS 1-5) for routine clinical diagnosis guidance. However, there is no consensus on whether PI-RADS 3 patients should go through biopsies. Mining features from these hard samples (HS) is meaningful for physicians to achieve accurate diagnoses. Currently, the mining of HS biomarkers is insufficient, and the effectiveness and robustness of HS biomarkers for prostate cancer diagnosis have not been explored. In this study, biomarkers from different data distributions are constructed. Results show that HS biomarkers can achieve better performances in different data distributions.

## Introduction

Prostate cancer is the most common type of solid organ malignancy in men worldwide[1]. Clinical studies have shown that prostate diseases present a differentiated clinical state from inert to highly aggressive[2]. Therefore, noninvasive and accurate diagnosis of clinically significant prostate cancer (csPCa) patients is very important, which can reduce excessive biopsy. Multi-parametric magnetic resonance imaging (mp-MRI) has already become an important diagnostic technique for prostate diseases. American College of Radiology and European Society of Urogenital Radiology proposed Prostate Imaging Reporting and Data System (PI-RADS) based on mp-MRI, which classified patients into 5 categories (PI-RADS 1-5) to aid in the diagnosis of csPCa according to the degree of malignancy with 1 being the lowest and 5 the highest [3-5]. However, severe controversies exist for the diagnosis of PI-RADS 3 patients regarding the necessity for them to go for biopsy [6,7]. Accordingly, these patients are treated as hard samples (HS), and more attention should be paid to these samples. Research in computer vision shows that putting more weight on the information extracted from difficult samples can help to improve the performance of the overall model [8-10]. Inspired by these successes, we believe that enhancing the information mining from mp-MRI of HS of prostate patients (PI-RADS 3 patients) can improve prostate cancer diagnostic performance. Nevertheless, most existing relevant studies perform prostate mp-MRI data mining from all collected samples without preference [11-13]. The biomarkers related to csPCa diagnosis contained in HS are not constructed properly, and the effectiveness and robustness of HS biomarkers have not been explored [14,15]. To this end, this study explores the effectiveness and robustness of radiomic biomarkers built with mp-MRI data of PI-RADS 3 patients for the diagnosis of csPCa. Experiments are conducted with data from three independent cohorts containing different distributions of samples, and results verify the effectiveness of the proposed method.

## Methodology

Detailed information of the experimental data utilized in this study is given in Figure 1. Among the three retrospective datasets, 204 patients in PD cohort are from the public data set of prostatex (Radboud University Medical Center), and the other two are from two medical centers in China, 574 patients in WH cohort (Wuhan University People's Hospital) and 51 patients in GD cohort (Guangdong Provincial People's Hospital). The PI-RADS evaluation of GD cohort and the segmentation of prostate in bp-MRI data of all patients were performed by two experienced doctors, and all results were examined by a senior expert. The proposed experimental procedure for obtaining the radiomic biomarkers is shown in Figure 2. A large number of quantitative radiomic features are obtained from the segmented prostate tissue. The least absolute shrinkage and selection operator (lasso) method with cross validation is used to perform feature selection and model construction. Three sets of radiomic biomarkers are obtained from the three cohorts with different prostate cancer patient distributions.

Training and validation datasets in each cohort are partitioned with a ratio of 7: 3 based on patient visit time. The training dataset is used to construct the csPCa prediction model. In this study, three csPCa diagnostic models are constructed with the same training dataset using the three sets of radiomic biomarkers (extracted from the three cohorts). The diagnostic performance and robustness of the constructed models are validated on the validation datasets from all three cohorts. The performance of models is quantified by the area value under the curve. Delong tests are performed to check the significance of the performance difference between different models. The value of the information contained in the three sets of radiomic biomarkers is analyzed for prostate disease diagnosis.

## Results and Discussion

In total, 1576 features are extracted from bp-MRI of each patient, and three sets of radiomic biomarkers are obtained from the three cohorts (GD/WH/PD) by filtering these features. Here, the biomarkers of GD cohort are the biomarkers constructed from only HS (PI-RADS 3 patients), whereas for WH and PD, patients of other PI-RADS scores are also included in the biomarker construction process. Experimental results are shown in Figures 3-5. It can be observed that csPCa diagnostic models based on GD biomarkers have better diagnostic performances on all three cohorts, validating the effectiveness and robustness of HS biomarkers. We speculate that the radiomic biomarkers obtained from equivocal PI-RADS 3 patients can better capture the image representation differences between csPCa and inert prostate diseases. Therefore, we suggest that future csPCa diagnostic studies may pay more attention to image data mining of PI-RADS 3 patients.

## Conclusion
In this study, we found that the radiomic biomarkers obtained from PI-RADS 3 patients have better diagnostic value in the identification of csPCa. In future research on MRI-based diagnosis of csPCa, it is recommended to consider strengthening the data mining of PI-RADS 3 patients.

## Acknowledgements
This research was partly supported by theNational Natural Science Foundation of China (61871371, 81830056, 81801691,61671441), Youth Innovation Promotion Association Program of Chinese Academy ofSciences (2019351), Collaborative Innovation Major Project of Zhengzhou (20XTZX06013).

# Figures

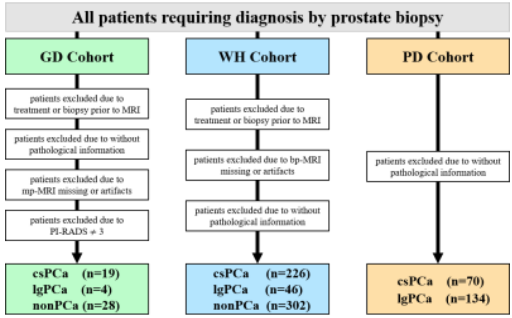

Figure.1 Flowchart shows patient selection process. bp = Biparametric, mp = Multiparametric, csPCa = clinically significant prostate cancer, lgPCa = low grad prostate cancer, nonPCa = Other diseases other than cancer.

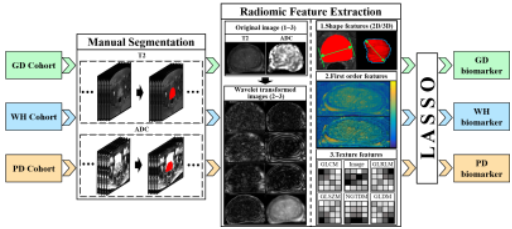

Figure.2 Biparametric MRI-based radiomic biomarker acquisition flowchart. Quantitative features are extracted from the tumor areas in brain MR images. The most important features are selected to train the LRC. The outputs of the trained LRC are treated as the radiomic signature that contain expected human knowledge.

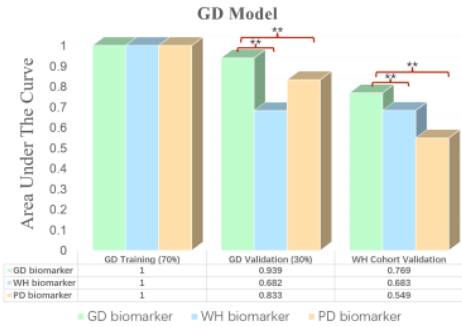

Figure.3 The prediction model of csPCa was constructed based on different Biomarker using GD Cohort training data. Two asterisks indicate significant differences between the two results in the Delong test.

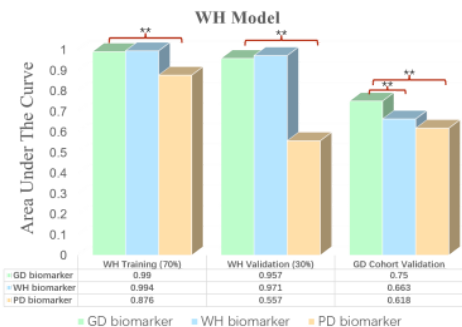

Figure.4 The prediction model of csPCa was constructed based on different Biomarker using WH Cohort training data. Two asterisks indicate significant differences between the two results in the Delong test.

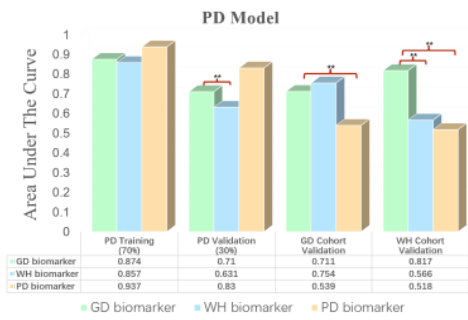

Figure.5 The prediction model of csPCa was constructed based on different Biomarker using PD Cohort training data. Two asterisks indicate significant differences between the two results in the Delong test.